# Biosphere Substrate and its parameters range

Yegor A. Morozov*[ab], Mikhail Bukhtoyarov[c], Mahdi Yoozbashizadeh[a]

* Expansist@gmail.com

[a] Mechanical & Aerospace Engineering Department, California State University Long Beach, 1250 Bellflower Blvd.,, Long Beach, California, 90840, USA

[b] Institute for Mathematical Sciences, Claremont Graduate University, 150 E. 10th Street, Claremont, California, 91711, USA

[c] Faculty of Liberal Arts and Sciences, I St, Budva, 85310, Montenegro

## Highlights

- Biosphere Substrate, celestial body suitable for Terraforming & inhabitation by our posterity, was defined
- Biosphere Substrate parameters ranges were determined, not only instant but over Planetary System lifecycle
- This is initial estimate which celestial bodies are economically worth Terraforming

## Abstract

For providing the maximal carrying capacity and expansive potential of a Planetary System, optimal utilization of the orbital material is required. It raises the question of suitability of various objects in a planetary system for the purpose of terraforming and developing full-scale open biospheres.

The term "Biosphere Substrate" is introduced for celestial bodies, like terrestrial planets or big gas giant moons, suitable to sustain a full-scale open biosphere after Terraforming. The purpose of the work is to examine the range of parameters for a biosphere substrate.

Most importantly, to be a Biosphere Substrate, a celestial body gravity should sustain a full-scale open breathable atmosphere dense enough after Terraforming in astrophysical time scale comparable to the lifecycle of a Planetary System. Humans can't reproduce in too low gravity, and can't restore functionality after long exposure to microgravity. Extremely high gravity is also the limit for our posterity forms functioning. The gravity is the main parameter that is the most difficult to change, thus the range of celestial object sizes fitting the Biosphere Substrate definition is examined first.

The capability of a celestial body to retain breathable atmosphere during a time interval comparable to the planetary system life cycle depends on its magnetic field strength. The lower boundary of the objects' sizes fitting the Biosphere Substrate definition is studied on the basis of their capability to hold hot convecting interiors for strong enough magnetic fields during a Planetary System life cycle.

As the chemical composition of a Biosphere Substrate's surface, hydrosphere and atmosphere can be drastically adjusted in most cases by importing asteroid and cometary material, this factor in most cases does not determine whether a celestial body can become a Biosphere Substrate. But for water worlds where any solid material to build biological bodies and industrial structures can be more than thousand kilometers deep, chemical composition can become an issue, while this problem can be solved to some extent by importing solid materials (asteroids). Also, concentrations of some chemical elements and compounds in extremely toxic quantities in the crust might be an obstacle to develop a full-scale open biosphere. Thus, the range of chemical compositions for celestial bodies of suitable sizes is also reviewed.

The temperature is another important parameter, as a molten or too hot celestial body cannot become a Biosphere Substrate in near term, but can become a Biosphere Substrate when it cools enough. The temperature factor and capabilities to adjust it are also estimated.

For a full-scale open biosphere development, there must be a sufficient source of power. A Biosphere Substrate orbit can go inside the circumstellar habitable zone, or can be partially inside and partially outside it for highly elliptical orbits. Possibilities to extend a circumstellar habitable zone with large orbital light diffusers and light concentrators are researched.

For planets with solid crusts, their rotation can be adjusted by hitting asteroids and comets strictly tangentially with high velocities to gain required spin. But for water worlds and tidally locked planets, rotation can be an issue. The rotation factor is also examined.

The work attempts to introduce a new concept and the term that can be used for further terraforming research, developing a mathematical model of its limits. Our estimations are required to be discussed with a wide range of specialists in the relevant fields of studies, such as astrophysics, especially exoplanets characterization professionals; astrochemistry; astrobiology; planetary science, etc. To provide the basis for this discussion, the article suggests preliminary estimations of the parameters crucial for selection of Biosphere Substrate candidates.

**Key words:** terraforming, biosphere substrate, planetary habitability, gravity, pressure, circumstellar habitable zone, planetary orbits, irradiation, temperature, Shelford's bell curve,

## 1. Introduction

The purpose of this work is to define the parameters range of Biosphere Substrate - celestial bodies suitable for Terraforming & inhabitation by our posterity in generations.

For the purpose of Terraforming [Birch, 1991, 1992; Zubrin, 1993; Visysphere Mars, 2005; Hossain et al, 2015; Brandon, 2018], to define the Biosphere Substrate concept, we are interested in Planetary Habitability for our posterity [Cockell et al., 2016], which is the number of humans a Planet can feed & provide environmental conditions for reproduction & workplace for further Cosmic Expansion of future generations. Full scale planetary uncontained open Biosphere as the result of Terraforming - is evolving biogeocoenosis [Smil, 2003; Dietrich et al, 2006; Rispoli, 2014] covering all surface area of the Biosphere Substrate with dense photosynthesising layer without need for greenhouses or other shelter, like we are approaching on Earth lately. Almost every dependence in Biology is a bell curve - maximum productivity for life is somewhere in-between, while both extremes, too much or too few of something, are deadly for life, as illustrated in example on Fig. 1. Planetary Habitability [Güdel et al, 2014; Vladilo et al, 2013] is evolving over time, can be both increased (Terraforming, Expansion) or decreased (destruction, ecocide). Not only instantaneous habitability might matter - but also conditions for life to emerge & develop. Continuous planetary habitability duration matters, because production is the workforce (number of workers) multiplied by time they work. Terraforming is the process of technically extending Planetary Habitability for our posterity, so the Planet can feed & support more humans over more time. Earth might be even not the most habitable planet - superhabitability [Heller & Armstrong, 2014, Heller 2015] is possible, Super-Earths can feed more population that Earth size or smaller Planets, and during longer time periods remain habitable in same orbits due to lower atmospheric escape thanks to bigger mass & stronger magnetic field retaining atmospheres longer.

Biosphere Substrates are mostly Terrestrial Planets & smaller edge of gaseous or icy planets, which includes [PHL, 2018; Hill et al, 2023] Subterrans (Mars size), Terrans (Earth size), Superterrans (Super-Earths & Mini-Neptunes), & in some cases Neptunians (lower range of Neptune size), but may also might include some biggest moons of Jovian planets, whose size & gravity is comparable to Terrestrial Planets. Biosphere Substrates are always determined in context of Planetary Systems they are in, depending on material available for Terraforming and to what distances can we beam sunlight power to deliver that material [Morozov et al, 2021]. This includes technically extended long-term Habitability during the Planetary System lifecycle, for a Planetary System Configuration process optimisation, in order to determine Biosphere Substrates - Terraforming Candidates that can be technically and economically made habitable with realistic means during time scales of an order of magnitude comparable to the particular Planetary System lifecycle, which means similar order of magnitude, or no less than 10% of the Planetary System lifecycle.

Circumstellar Habitable Zone (CSHZ) is the range of orbits on which liquid water can exist. Technically Extended Circumstellar Habitable Zone - examines technical limits of extending circumstellar habitable zone; currently limited by capabilities of large orbital starlight energy dissipators and concentrators.

Most works on terraforming [Fogg, 1998; Graham, 2004] discuss only biological means, but Orbital Engineering to deliver the limiting chemical elements by Planetesimals Redirection operations for Terraforming is required prior to biological seeding.

Planetary Systems Configuration - process of tugging and colliding asteroids, comets, maybe dwarf planets and moons (planetesimals in general) in order to achieve several Biosphere Substrates (biosphere substrate - a terrestrial planet or a gas giant moon, planned to host full-scale open biosphere) with chemical compounds, orbital and physical characteristics, etc., as close to optimal (by criteria of maximal overall carrying capacity and expansive potential) as possible, that are inside the star (or double-triple system) Circumstellar Habitable Zone (CSHZ) during most of Planetary System lifecycle determined by its host star evolution.

Planetary Habitability ranges differ significantly for complex multicellular life like humans, and for primitive unicellular life [Lammer et al, 2009] that is not capable to Terraform & inhabit other Biosphere Substrates, thus will be extinct due to stellar life cycle if doesn't manage to develop full scale open biosphere with technologically capable species like humans to Terraform & inhabit other Biosphere Substrates soon enough [Morozov et al, 2021].

Planetary System Configuration and Terraforming processes (hundreds to thousands years) [Hossain et al, 2015; Brandon, 2018] take many (about 6-9) orders of magnitude less time scale than a Planetary System life cycle (about billions years for G-type stars and hundred billion years for M-type stars). Thus, the small time scale of thousands years work for optimal Planetary System Configuration processes and all the Terraforming efforts must be taken very meticulously, as it drastically determines the carrying capacity thus population thus workforce of a Planetary System during billions of years time scale of its lifecycle, certain Biosphere Substrates lifespan and maintenance cost on different evolutionary stages, and the Planetary System total productivity for further Cosmic Expansion of our posterity and life as whole.

Planetary Habitability evaluation requires consideration of not only the Planet's orbit semi-major axis, but also many more factors [Zsom, 2015]. These factors may include the planetary rotation including daylength & tilt, with consequences on climate and magnetic field generation [Blanc et al, 2005], the relationships between the Planet mass/radius and the atmosphere and plate tectonics, the role of volatiles in the hydrosphere, atmosphere and plate tectonics, the atmosphere evolution and escape rates, and the co-evolution of an atmosphere and hydrosphere. The effects of all these geodynamic processes on Planetary Habitability and ultimately on the reproduction and development of our posterity should be also considered.

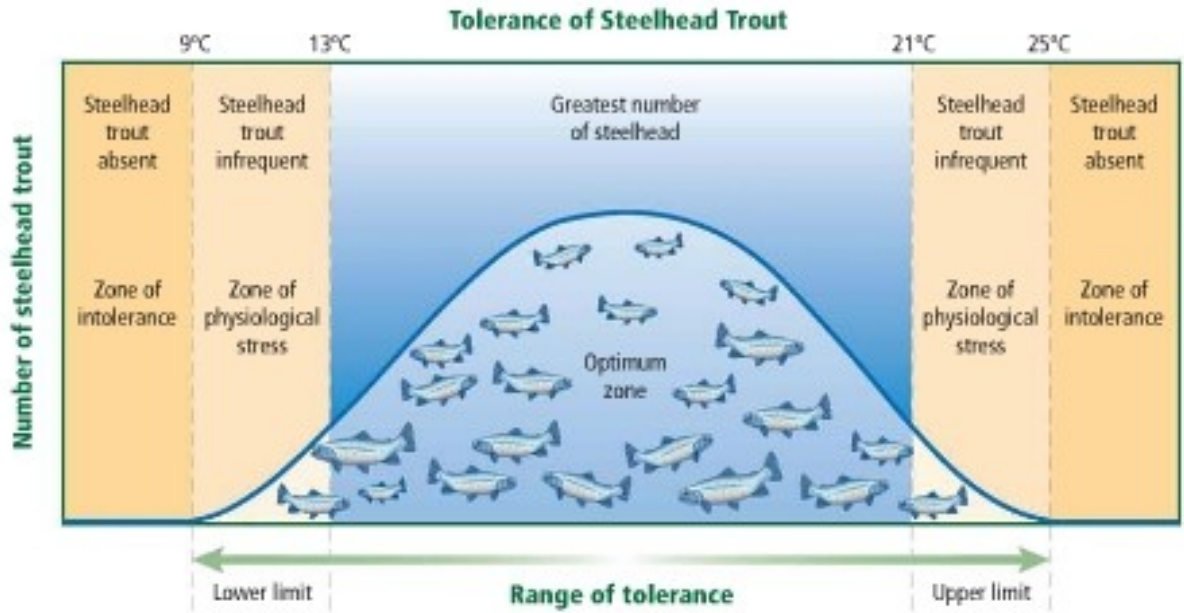

Fig. 1. Shelford's Bell Curve (Shelford's law of tolerance)

## 2. Material and methods

This study evaluates the physical and environmental parameters that define a Biosphere Substrate — a celestial body capable of sustaining a full-scale open biosphere after Terraforming for our posterity to reproduce in, during sufficiently long timescales to pay off Terraforming efforts. The approach integrates astrophysical models, planetary science, geophysics, biological tolerance studies, and conceptual space engineering Terraforming methods.

### *2.1 Assumed Terraforming Technologies*

Assumed Terraforming Technologies include most importantly tugging Planetesimals [Zubrin et al, 1993] with ion thrusters using evaporated Planetesimal material as propellant as-is powered by concentrated beamed solar power [Morozov et al, 2021; Morozov et al, 2025], and orbital sunlight concentrators [Forward, 1984; Frisbee, 2004; Birch, 1992] and orbital sunlight dissipators [Birch, 1991].

The biggest, most important, most expensive work in Terraforming, to most extent determining planetary carrying capacity thus population thus workforce over the remaining Planetary System lifecycle, is delivering the limiting chemical elements by Solar-powered Planetesimals Redirection for Terraforming operations, because the amount of biological cells and organisms that can assemble from a Biosphere Substrate, is always determined by the limiting chemical element in the biological stoichiometric proportion [Morozov et al, 2021; Morozov et al, 2025]. Planetesimals on outer orbits of Planetary Systems are abundant and diverse in chemical composition, from number depending on size proportion there are assumed to be millions of at least km size Planetesimals in Kuiper Belt and even much more beyond in Oort Cloud, with total mass of those millions of Planetesimals bigger than mass of a terrestrial planet, allowing to assemble optimal surface chemical composition including atmosphere + hydrosphere + soil, on all the Biosphere Substrates in our Planetary System [Bockelée-Morvan et al, 2004; Mumma et al, 2011; Vilenius et al, 2014; Fernández, 2020], and similar situation is observed in ExoPlanetary Systems [Deeg et al, 2018]. Solar-powered Planetesimals Redirection for Terraforming also allows to neutralize toxic compounds like $SO_2$ in Venus atmosphere by spraying metal asteroids in upper layer of atmosphere [Birch, 1991], strengthen planetary magnetic fields by assembling moons in significantly eccentric orbits from least valuable leftover Planetesimal material [Pearson, 2014; Hossain et al, 2015], adjust planetary rotation including daylength and tilt which is very important for Terraforming planets like Venus [Morozov et al, 2021; Morozov et al, 2025]. As most Biosphere Substrates surface chemical composition is far from optimal for life, Solar-powered Planetesimals Redirection for Terraforming to deliver the limiting chemical elements, optimize planetary rotation, and strengthen planetary magnetic fields - is the main most important ubiquitously universally necessary work in almost any Planetary System. The more distant from the host star - the more rich in volatiles, abundant and numerous Planetesimals are [Morozov et al, 2021; Morozov et al, 2025], and smaller Planetesimals are more rich in volatiles thus more valuable for importing atmosphere & hydrosphere for Biosphere Substrates, while moons can be assembled from biggest Planetesimals that are more rocky. Tugging asteroids, comets, and other planetesimal material from beyond Neptune to Mars, Venus, and other Biosphere Substrates, by Ion Thrusters using evaporated planetesimal material as propellant as-is, powered by concentrated precisely beamed starlight power, is possible through entire Kuiper Belt with presently existing materials & technologies, and delivering Planetesimals from Oort Cloud might be possible with 1-3+ orders of magnitude principal pointing & tracking angular precision increase [Morozov et al, 2025]. Solar-powered Planetesimals Redirection for Terraforming concept includes thousands of Planetesimal Tug Spacecraft Systems (PTSS) that are spider-like robots consisting of mainly 2-axis gimballed photovoltaic arrays powering 2-axis gimballed ion thrusters arrays that soft land on each target Planetesimals and deliver it to the target Biosphere Substrate, and Power Harvesting & Beaming Systems (PHBS) that are ultra-thin ultra-lightweight orbital Fresnel lenses with 2-axis gimballed flexible optic fiber collimators ended by solar-pumped laser to decrease beam divergence, tracking each PTSS photoreceivers to power it during each entire mission [Morozov et al, 2021; Morozov et al, 2025]. Yunitsky's General Planetary Vehicle Orbital Launch Ring exactly on all Equator of every Terraformed & inhabited Biosphere Substrate is necessary for cheap large-scale cargo flow from surface to orbit [Yunitskiy, 1982, 1987] for Solar-powered Planetesimals Redirection for Terraforming [Morozov et al, 2021; Morozov et al, 2025].

  Active magnetic field strengthening methods such as starlight-powered solenoids along Equator for strengthening planetary magnetic fields can be used if passive methods such as assembling a moon in eccentric orbit to create tidal forces thus strengthening geodynamo + heating the planet including interiors by planetesimals impacts + creating inhomogeneity achieved by Planetesimals Redirection operations for Terraforming effects were not enough [Morozov et al, 2025] - but active methods to strengthen planetary magnetic fields should be used only after passive methods have been completely implemented and their potential is exhausted proven insufficient, because active methods of planetary magnetic fields generation take significant fraction of most valuable Equatorial latitudes surface area that can be used for photosynthesis feeding more population thus workforce otherwise, for photovoltaic arrays to power them, and required constant maintenance over the remaining planetary lifecycle.

Ultra-thin ultra-lightweight nano-foil orbital starlight dissipators for inferior Biosphere Substrates like Venus [Birch, 1991], and ultra-thin ultra-lightweight nano-foil orbital starlight concentrators for exterior Biosphere Substrates like Mars and further [Birch, 1992; Forward, 1984; Frisbee, 2004] can significantly artificially Technically Extend natural circumstellar habitable zones both inwards and outwards, making more Planets into to Biosphere Substrates after completing the Terraforming works - yet first Planet Terraformed inside a new Planetary System should be inside natural CircumStellar Habitable Zone because it is very difficult & expensive to transport planetary-scale sunlight concentrators or dissipators on a starship, so they should be made from first Terraformed and inhabited Biosphere Substrate material in a new Planetary System for other Biosphere Substrates [Morozov et al, 2025].

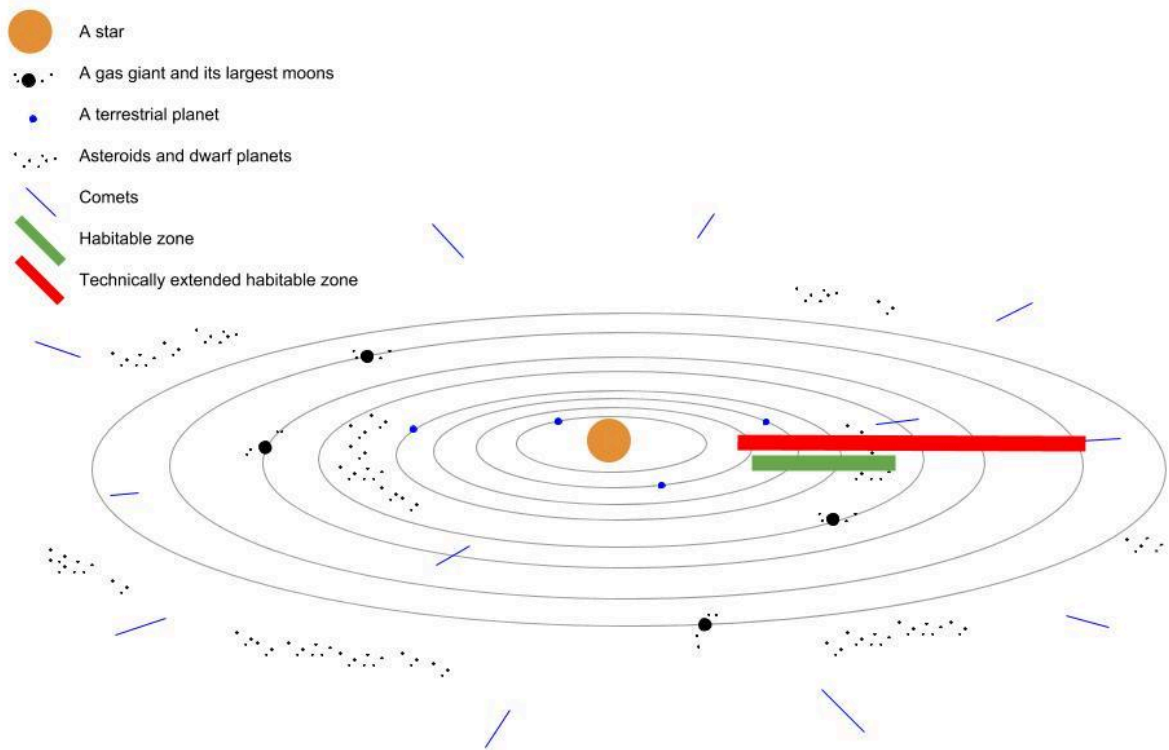

Fig. 2. Technically Extended Circumstellar Habitable Zone [Morozov et al, 2018]

Biological seeding is the easiest and cheapest part of Terraforming, well described in many earlier publications as gradually introducing organisms from simplest to more complex to achieve forest habitability: starting with unicellular photosynthesising organisms to build initial organic material and oxygen-rich atmosphere; after initial organic material built, seed grass &

legumes to build initial solid; after initial soil built - introduce fruit threes & all kinds of edible plants, after food grow - introduce animals & humans, Terraforming complete [Graham, 2004].

## *2.2 Methodological Approach of parameters ranges determination*

Based on Planetary Habitability factors reviews [Cockel et al, 2016; Brandtl, 2015; Lammer et al, 2009; Güdel et al, 2014; Rodríguez et al, 2017], we determined main parameters for a Biosphere Substrate concept, for which we estimate lower and upper thresholds:
1. Gravity (g) range, because it determines human reproduction and operation, and thermal atmospheric escape rates.
2. Temperature (K) (average planetary) range, because a rather narrow temperature range allows complex life to exist.
3. Starlight Irradiation (W/m$^2$) range because only photosynthesis can power large-scale long-lasting Biospheres development as any supplies will deplete fast on astronomical timescales.
4. Atmospheric Pressure (atm) range including partial pressures of certain gasses, because at too low or too high pressure humans die, and needed for hydrosphere and photosynthesis.
5. Planetary Magnetic field strength (T) range, because it is necessary for humans & other animals reproduction bioregulation [Binghi, 2002], breathable atmosphere retention time increase, and protection against penetrating radiation.
6. Breathable Atmosphere lifetime (Myr) above certain level to pay off Terraforming costs.

Biological thresholds were identified from experimental studies with humans or animals similar to humans if no data on humans available, and theoretical analysis based on them, including minimum and maximum tolerances of humans or other mammals for ranges of gravity, atmospheric pressure, temperature, and more.

Atmospheric escape models were applied to estimate the timescales for breathable atmosphere retention, using both thermal and non-thermal escape processes as functions of planetary mass, radius, and magnetic field strength.

Orbital habitability evolution was examined in the context of CSHZ natural evolution, with consideration of physically reasonable technically extended CSHZ methods.

Parameter synthesis combined these factors into quantitative ranges to determine viable Biosphere Substrate candidates.

## *2.3 Atmosphere escape rates determination*

Atmosphere Escape Rates (AER) determination and Breathable Atmosphere Lifetime (BAL). Atmospheric loss/erosion depends mainly on Biosphere Substrates' gravity, temperature and magnetic field strength.

Many different atmospheric escape mechanisms take place simultaneously, we are most interested about the limiting factors of viable breathable atmosphere decay time.

For thermal atmospheric escape, atmosphere half-life is linearly directly proportionate to the celestial body mass and inverse proportionate to its radius [Strobel, 2009]:

$$\lambda = \frac{G*M*m}{r*k*T_0} \qquad (1)$$

where **G** is Newton's universal gravitational constant, **M** is the mass of the Biosphere Substrate, **r** is the radius of the Biosphere Substrate, **m** is the mass of the dominant molecule (N$_2$ for example), **k** is Boltzmann's constant, **T** is the atmospheric temperature, **v** is the bulk outflow velocity, **c** is the isothermal speed of sound = $(k*T_0/m)^{1/2}$, $c_p$ is the specific heat at constant pressure, $k = k_0*T$ is the thermal heat conductivity with $\kappa_0 = 9.37$ erg*cm$^{-1}$*s$^{-1}$*K$^{-2}$, Q is the net heating/cooling rate per unit volume, F is the outflow rate in molecules s$^{-1}$*sterad$^{-1}$ and subscript "0" denotes the lower boundary. [Strobel, 2009].

For non-thermal atmospheric escape [Luo et al, 2023], there are several major mechanisms known [Hunten, 1982]:

    1. Charge exchange
    2. Dissociative recombination
    3. Impact dissociation and photodissociation
    4. Ion-Neutral reactions
    5. Sputtering of knock-on
    6. Solar-wind pickup
    7. Ion escape
    8. Electric fields

Generally, all the planetary non-thermal atmospheric escape rate factors are significantly mitigated by strengthening the planetary magnetic field [Chassefière et al, 2007; Luhmann et al, 1992; Lammer & Bauer, 1991], and highly nonlinearly depend on the XUV power fraction in its host star's radiation. According to [Gallet et al., 2016] the relation between the minimum planetary magnetic field, required for an effective magnetic protection, and the stellar magnetic field is:

$$B_p^{min} \simeq 16052 * \left[ \frac{B_*^2}{8\pi} * \left( \frac{2.5*R_*}{1AU} \right)^4 \right]^{1/2}$$

(2)

According to [Forget & Leconte, 2014], one of the XUV atmospheric escape dependences is:

$$F_{esc} = \eta \frac{R_P F_{XUV}}{GM_P} (\text{kg m}^{-2}\text{ s}^{-1}),$$

(3)

where $F_{XUV}$ is the averaged XUV flux received by the planet (i.e. divided by 4 compared with the flux at the substellar point), G is the universal gravitational constant, and $R_p$ and $M_p$ are the planetary radius and mass, respectively.

## *2.4 Assumptions and Limitations*

This work assumes advanced Terraforming technologies including orbital engineering and the limiting chemical elements delivery to Biosphere Substrates by Planetesimal redirection operations. Chemical composition is treated as a secondary factor, as it can often be modified through external material delivery, except in cases of extreme toxicity or water worlds with inaccessible solid material or gas giants with too high pressure and gravity. Biological productivity is modeled under the principle of Shelford's law of tolerance, assuming optimal productivity occurs between extremes, closer to mid-range conditions of planetary parameters.

## 3. Theory & calculations

Almost every dependence is biology is a bell curve, with maximal productivity for life in-between and both extremes deadly. Thus, for every important planetary parameter we must define the acceptable range of minimum and maximum values.

### *3.1 Gravity Constraints*

Minimum Planetary gravity. It is known that microgravity [Rydze et al, 2017] is harmful for mammal reproduction, functionality, and long-term population survival [Ruden et al, 2018; Narayanan, 2023]. Humans can't restore their functionality after more than 1-2 years in microgravity [https://ntrs.nasa.gov/api/citations/20160001730/downloads/20160001730.pdf] and can't reproduce in microgravity [Miglietta et al, 2023]. But how much, what fraction of Earth gravity can be considered sufficient for our posterity long-term normal functioning & reproduction during many generations?

Research on the effects of partial gravity on reproduction indicates that the threshold for successful reproduction is probably species-dependent, with invertebrates demonstrating far greater tolerance to microgravity than vertebrates [Clément et al, 2006]. Fruit flies can complete reproductive cycles even in microgravity, though alterations in gene expression and stress responses have been observed [Iyer et al, 2022]. Studies have found that microgravity conditions do not stop the reproduction of fish and amphibians [Gualandris-Parisot et al 2002; Murata et al 2015; Aimar et al 2002; Wolgemuth 1995]. Fish such as medaka can undergo fertilization and early embryogenesis in orbit [Wolgemuth, 1995]. In contrast, mammalian reproduction appears more dependent on stronger gravitational cues. While extensive data details the profound disruptions caused by microgravity on mammalian gametogenesis, early embryogenesis [Ruden et al., 2018; Cheng et al., 2023; Li et al., 2021], and ovarian function [Kikina et al., 2024], and hypergravity acutely impacts sperm motility [Ogneva et al., 2024], the reproductive consequences of partial gravity (e.g., Martian or Lunar levels) are critically understudied. Comparative analyses highlight significant species-specific vulnerabilities; mammals exhibit substantial sensitivity, evidenced by altered fetal positioning and parturition even after gestation in microgravity [Ronca & Alberts, 2000]. Crucially, direct experimental data defining the *threshold* of gravity necessary for viable mammalian multi-generational reproduction is absent. [Donovan et al. 2018] explicitly identify this gap, noting the paucity of studies on rodent responses to partial gravity simulations and advocating for targeted research. Proposed mission concepts like MICEHAB [Rodgers & Simon, 2015; Rodgers et al., 2016] aim to address this by studying autonomous rodent colonies in partial gravity, recognizing that determining whether gravity levels below 1 g but above microgravity (e.g., 0.16 g, 0.38 g) are sufficient to support normal fertilization, embryonic development, gestation, and birth across generations. Human reproduction at different levels of partial gravity have not been studied well enough yet. Given the severe impacts observed in microgravity and the lack of data, mammalian reproduction likely requires gravity levels substantially higher than microgravity, potentially near Martian gravity (0.38g) [Richter et al, 2017]. However, the precise minimum threshold remains unknown and represents a critical knowledge gap for long-term space habitation [Richter et al, 2017]. Yet, there was some preliminary research at NASA Johnson Space Center on human functionality in partial gravity, suggesting **0.3 g** is the bare minimum required for long-term human functionality [Clement et al, 2015.]. Artificial Gravity Animal Research prior to ISS Utilization [Soviet space research 1961] - testing rats and mice in reduced gravity during parabolic flight. The posture and locomotion of the animals appeared normal during brief periods of **0.3 g** exposure, thus setting this as a minimum gravity level requirement for locomotion [Yuganov et al. 1962; 1964]. The first animals to be centrifuged in space were flown on the 20-day Cosmos-782 mission in 1975, when fish and turtles housed in containers were centrifuged at 1 g. In 1977, a significantly more extensive investigation was executed using rats that were centrifuged during the 19-day mission of Cosmos-936. Based on long-duration centrifuge studies on Earth, Russian scientists suggest that the bare minimum level of effective AG in humans is about **0.3 g**. They further recommend that a level of 0.5 g be induced to increase a feeling of wellbeing and normal performance [Shipov et al. 1981]. Consequently the lower limit of 0.3 g was preferred in most design studies for implementation of artificial gravity [Letko & Spady 1970].

Maximum Planetary gravity. Regarding the highest limit of Biosphere Substrate gravity, it has been shown that hypergravity levels of 2 g have positive strengthening effects for mammal development, while 3 g causes deleterious effects on mammals development & reproduction [Gnyubkin et al, 2015], thus we might assume that mammals can reproduce over generations at up to 3 g environment certainly, but **3 g** hypergravity is significant stress near the upper limit of mammals tolerance. Again, we lack experiments about human functioning & reproduction at hypergravity of 2g, 3 g, etc., so we will base our assumptions on experiments with mammals closest most similar to humans that such experiments were conducted on. Thus, we will take **3 g** as the first estimate of the upper limit of Biosphere Substrate gravity, which can be increased or changed when more experimental data becomes available.

### *3.2 Temperature Constraints*

Temperatures range. Maximal temperature under which humans can operate is fixed if not taking into account long evolution where organisms develop significant adaptations to changing environments. Any complex life has relatively narrow temperature range constraints of **273.15 K < $T_{BS}$ < 323.15 K** [Silva et al., 2017]. These are constraints for mean average planetary temperature, because on Earth we see many life forms including humans survive short-term air temperature drops below **223 K**. Temporary planetary temperatures below water freezing point can be survived with clothers & housing. We assume that we can concentrate or dissipate starlight with large orbital lenses in a wider range of orbits than the natural Circumstellar Habitable Zone.

### *3.3 Irradiation range*

Irradiation range. The more Photosynthetically Active Radiation (PAR) per unit of surface area - the more productive photosynthesis this planetary ecosystems and industries, within the temperature frame humans can tolerate. Thus, PAR should be maximized under a certain maximal temperature tolerated by humans on all Biosphere Substrates. Thus, we should keep greenhouse gases low in any Biosphere Substrates atmospheres, to have higher PAR irradiation within tolerable temperatures. $CO_2$ concentration should be optimised by the trade-off between photosynthesis efficiency and humans tolerance threshold, while $CH_4$ and other greenhouse gases not required for photosynthesis are required to be carefully cleaned from the atmosphere - fixed into solid or non-volatile liquid compounds. Possible climates on terrestrial exoplanets are discussed in [Forget & Leconte, 2014]. Photosynthesis has been observed starting around **0.01 W/m²** irradiation levels [Hoppe et al, 2024], though extremely inefficient, extremely low productivity, requiring extremely much surface area to feed a human. No certain upper irradiation level killing any life forms is known, clearly there are organisms that can survive

**1370+ W/m²**, but most Earth's photosynthesising organisms have photosynthesis inhibition around 1000 W/m² due to damage of the photosynthesising organs by long duration high irradiation levels [Lingwan et al, 2023], though some organisms are thought to be able to survive higher irradiation levels including flares [Mullan & Bais, 2018].

### *3.4 Pressure Thresholds*

Minimum atmospheric pressure. Absolute theoretical limit for certain death is around **0.0618 atm** as the Armstrong Limit [https://en.wikipedia.org/wiki/Armstrong_limit] when humans die from internal liquids boiling can be taken as the lowest theoretically possible boundary. But humans can die much earlier due to low partial pressure of oxygen in an atmosphere with the same relative gasses proportion as on Earth, so survival at low pressures requires an oxygen rich atmosphere. Some research on Terraforming suggests that >**0.13 atm** $O_2$ partial pressure might be the lowest limit to breath for human operation [Pazar, 2017]. Yet, some humans can have health issues even at higher atmospheric pressures [West, 1999]. Also at least above >**0.01 atm** $N_2$ partial pressure is required for plants to grow [Pazar, 2017].

Maximum atmospheric pressure. Might be about **2.5 atm** for humans for the same gasses proportion atmosphere as on Earth, **78% $N_2$ 21% $O_2$**, due to toxicity of oxygen high partial pressure. Yet, humans have survived for weeks at **60 atm** in an atmosphere with **0.8 % $O_2$** and the rest **He** and/or **$H_2$** for deep sea diving https://space.stackexchange.com/questions/10895/maximum-survivable-atmospheric-pressure . The rest 99.2 % of an atmosphere don't have to be He & $H_2$ only, they can include N at levels that don't cause nitrogen narcosis, other gasses at levels below toxic partial pressures, and other inert noble gasses. N is some of the most valuable elements as it is the main essential building block of all proteins and usually the limiting chemical element in Earth agriculture, so N should be preserved and converted into soil using nitrogen fixing bacteria & into fertilizers & other useful things, if it is present in an atmosphere above levels safe for humans. Higher atmospheric pressure generally increases Planetary Habitability, making CSHZ broader and bigger around the same star compared to Planets with lower atmospheric pressure [Vladilo et al, 2013]. Higher atmosphere pressure also helps to retain enough hydrosphere necessary to irrigate plants.

### *3.5 Atmospheric Retention and Escape*

Atmosphere Escape Rates (AER) and Breathable Atmosphere Lifetime (BAL).

Biospheres cannot proliferate efficiently without humans or other numerously populated technologically capable society as the organisation force, thus efficient biospheres proliferation in long-term space prospective means humans operate efficiently enough.

For thermal atmospheric escape, atmosphere half-life is linearly proportionate to the celestial body mass and inverse proportionate to its radius $\lambda = \frac{G*M*m}{r*k*T_0}$ [Strobel, 2009], mass grows approximately proportionate to the volume which is proportionate to the the 3rd power of the planetary radius, thus **M/r ≈ r³/r ≈ r²**, so if thermal atmospheric escape rate is much bigger than all other mechanisms of atmospheric escape rates combined - then atmospheric half-life is roughly proportionate to the Biosphere Substrate radius squared: **λ ≈ r²**, which means Super-Earth atmospheres survive much longer, and the bigger the Planet the better for atmosphere retention during the Planetary System lifecycle.

Non-thermal atmospheric escape mechanisms are multiple and complex, and different of these mechanisms dominate under different conditions [Hunten, 1982; Luhmann et al, 1992; Lammer & Bauer, 1991].

The borderline between rocky terrestrial planets and gas giant planets is considered to be around 1.8 Earth radius of the solid part [Lehmer et al 2017; Buchhave et al, 2014].

Thermal atmospheric escape pressure loss from 60 atm to 0,13 atm is ≈460 times decay.

$$\lambda = \frac{G*M*m}{r*k*T_0} = \frac{6.674\times10^{-11}[N\cdot kg^{-2}\cdot m^{2}]*M*4,65x10^{-26}kg}{r*1.3807x10^{-23}[J\cdot K^{-1}]*300K} = 7.5*10^{-12}[\frac{m}{kg}]*\frac{M}{r}$$

For Mars:
$$\lambda \approx 7.5*10^{-12}[\frac{m}{kg}]*\frac{6.4171\times10^{23}kg}{3.390\times10^{6}m} \approx 1,42*10^{6}$$

### *3.6 Geodynamics and Magnetic Fields*

Plate tectonics is an important factor for planetary habitability, because plate tectonics allows carbon cycle on geological timescales acting as thermostat decreasing temperature fluctuations allowing more temperate habitable climate, renews heavy elements on a planetary surface via volcanism, and provides long-term cooling of planetary core, which is vital for generating planetary magnetic field capable of shielding atmospheric volatiles from stellar wind & radiation induced atmospheric escape during the Planetary System lifecycle [Foley & Driscoll, 2016]. Thus, a Biosphere Substrate should have plate tectonics, which makes it more habitable and valuable for our posterity reproduction. Yet, some scientists argue that poorer habitability can be still possible without plate tectonics in some cases [Tosi et al, 2017].

Strong magnetic fields of planets may be effective in hindering the hydrodynamic escape of planetary atmospheres, especially most valuable hydrogen and nitrogen. In addition to the magnetic field, the closer to a star the more important role stellar wind and radiation pressure will play in ExoPlanets AER [Tian et al, 2005; Chassefière et al, 2007; Ramstad et al, 2018].

Mars atmospheric escape rates depending on magnetic field models confirm the crucial role of planetary magnetic field strength for atmosphere retention [Fang et al, 2017; Ramstad et al, 2018; Shizgal et al, 1996].

The general trend is that the stronger magnetic field a Planet has, the bigger it is, the cooler it is - the lower the AER during the Planetary System Lifecycle [Vladilo et al 2013; Forget & Leconte 2014; Lehmer et al 2017; Catling et al 2009; Keating et al 2012; Lammer 2018; Lillis et al, 2015; Catling et al, 2009, Lammer et al, 1991, 2011].

### *3.7 Orbital Mechanics and Circumstellar Habitable Zone evolution*

CircumStellar Habitable Zone evolution [Jiang et al, 2024; Rushby et al, 2013; Brandl, 2015; Safonova, 2016] is an important consideration factor in Terraforming, which determines the habitable lifetime of each Biosphere Substrate. Both natural and Technically Extendable CircumStellar Habitable Zones, evolve over time with both inner & outer boundaries moving outwards as the Planetary System host star ages, becoming bigger & hotter, closer to its death by explosion [Ramirez et al 2014; Heller 2015; Safronova et al 2016; Heller et al 2016; Truitt 2017, Waltham 2017]. Stellar wind and radiation pressure also strengthens during Planetary Systems lifecycles as the host star ages [Lammer et al 2011; Lammer et al 2012]. Strong Ultraviolet radiation can impose additional constraints on Circumstellar Habitable Zones [Oishi et al, 2016].

Higher Eccentricities and Obliquities can slightly increase the outer boundary of Circumstellar Habitable Zones compared to Planetes with same orbit semi-major axes orbiting almost in ecliptic plane in almost circular orbits [Armstrong et al, 2014; Vladilo et al, 2013].

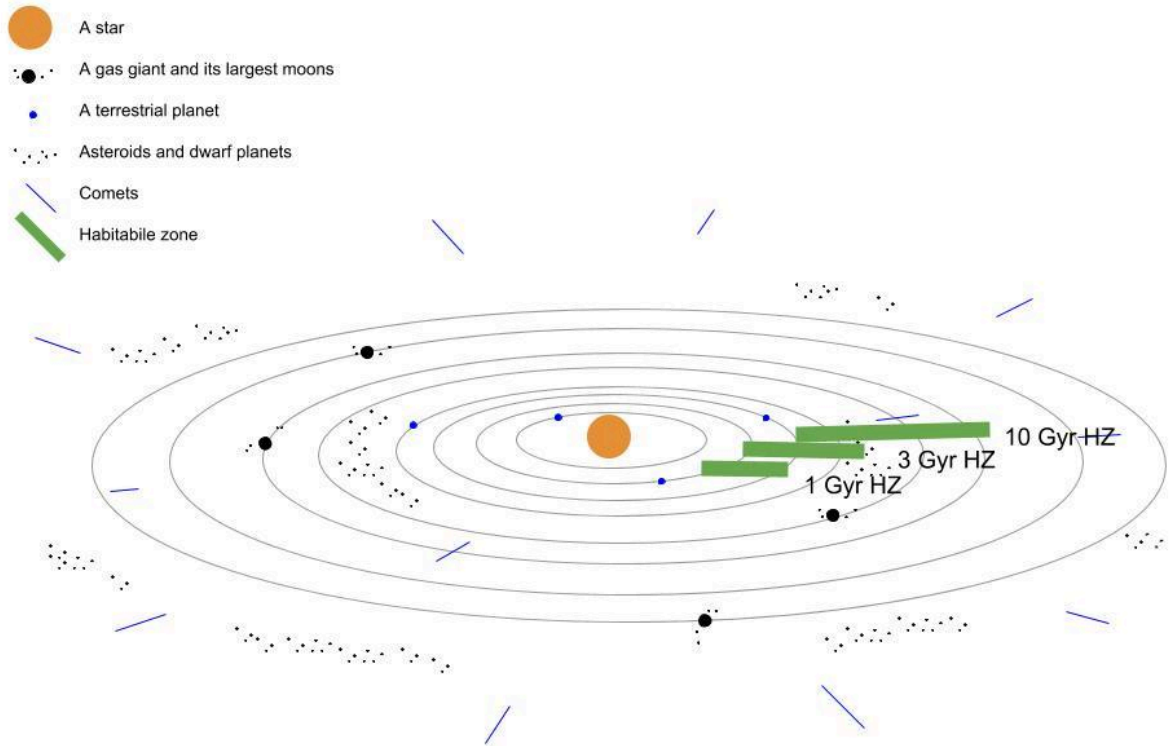

Fig. 3. CircumStellar Habitable Zone Evolution [Morozov et al, 2018]

Tidal locking can ruin planetary habitability - a satellite can increase time till a planet tidal locking dramatically, about Gyr [Barnes, 2017]. Planetary rotation including short enough daylength and tilt close to Earth's is important for planetary habitability to provide more even irradiation for photosynthesis and climate. Planetary rotation period thus daylength should be made smaller = planetary rotation angular rate should be set significantly higher than optimal during Planetary System Configuration & Terraforming, as planetary rotation is significantly slowing during Planetary System lifecycles. Earth day was about 21 hours - 600 Myr ago. Earth day was about 2-3 hours when the Moon was formed.

Principal physical limitations for technical extension of CSHZ:

Inferior - ability of an orbital light dissipator material to hold its shape mechanically and stay on the required orbit during the star's aging and luminosity increase is probably the major inferior technically extended CSHZ principal limit.

Exterior - size of an orbital light concentrator (Fresnel) lens till it's gravitational collapse and the lens and tracking pointing accuracy are probably the two major exterior technically extended CSHZ principal limits and the Planetary System Configuration processes range limits, determining the outer distance of moveable objects - asteroids, comets, and other planetesimals.

## 4. Results

This study establishes preliminary quantitative ranges for the parameters defining Biosphere Substrates — celestial bodies capable of supporting a full-scale open biosphere after terraforming. These ranges synthesize biological tolerance, astrophysics, planetary science, including atmospheric retention models.

| Parameter | Lower limit | Upper limit | Key Limitation |
|---|---|---|---|
| Gravity, g (*9.806m/s$^2$) | **~0.3 g** | **~3 g** | Reproduction, atmospheric retention |
| Temperature, average, K | **~273 K** | **~323 K** | Complex life tolerance |
| Irradiation, W/m$^2$ PAR | **>0.01 W/m$^2$** | **unknown** above >1370+ W/m$^2$ | Photosynthesis, photovoltaic power |
| Atmospheric Pressure, atm | **>0.13 atm** $O_2$ human breath <0.0618 atm certain death **>0.01 atm** $N_2$ for plants | 2.5 atm Earth proportion **60 atm** low-oxygen hydreliox | Armstrong limit, breath $O_2$ partial pressure, deep-sea diving records |
| Magnetic Field, T | Natural or engineered, strong enough to resist stellar wind erosion | None Observed | Atmospheric escape prevention, reproduction |
| Breathable Atmosphere Lifetime, Myr | **>1 Myr**, to make Terraforming efforts economically rentable | None, the longer the better | Terraforming efforts pay off, compared to other Terraforming targets |
| Orbit Semi-major Axis, AU | Orbital Sunlight Dissipator heat rejection to maintain mechanical properties | Orbital Sunlight Concentrators pointing & tracking precision limits | Technically Extended Circumstellar Habitable Zone; Stellar Lifecycle |

## 5. Discussion

Almost every dependence in biology is a bell curve: too high gravity, pressure, temperature, etc. is death, too low gravity, pressure, temperature, etc. is death - maximal life productivity is somewhere in-between.

The results presented here define preliminary thresholds for identifying Biosphere Substrates - celestial bodies capable of supporting full-scale, open biospheres after terraforming. These ranges, while provisional, provide a structured foundation for evaluating both Solar System bodies and exoplanetary candidates.

By quantifying the conditions under which a planet or moon may serve as a BS, this work introduces a conceptual tool for systematically prioritizing terraforming targets. Unlike existing habitability indices, which often emphasize conditions for microbial life, the BS framework is explicitly oriented toward sustaining complex ecosystems and human societies over astrophysical timescales.

Critical uncertainties remain in several areas:
- Reproduction in partial gravity [Rodgers & Simon, 2015; Richter et al., 2017; Mishra & Luderer., 2019; Sharma et al., 2024]: While mammalian reproduction fails in microgravity, the precise lower threshold for sustainable multi-generational reproduction remains unknown.
- Atmospheric escape dynamics: Non-thermal escape mechanisms are complex, highly dependent on stellar winds and magnetic fields, and not fully constrained by current models.
- Long-term magnetic field sustainability [Foley & Driscoll, 2016]: The evolution of planetary dynamos and their interaction with stellar activity remain major unknowns for atmospheric retention.

These gaps highlight the need for targeted experimental research, particularly in reproductive biology under partial gravity and planetary magnetohydrodynamics.

## 6. Conclusions

We have assessed Biosphere Substrates, celestial bodies suitable for Terraforming & inhabitation, parameters range. Biosphere Substrate parameters ranges:

Biosphere Substrate parameters range:
Gravity range within $0.3 \cdot g_{Earth} < g_{BS} < 3 \cdot g_{Earth}$
Temperature (average planetary) range within $273.15\ K < T_{BS} < 323.15\ K$
Irradiation range within, $W/m^2 > 0.01\ W/m^2$
Atmospheric Pressure range within $0.13\ atm < P_{BS} < 60\ atm$
Planetary Magnetic field strength range, within T
Breathable Atmosphere lifetime within, Myr

**Acknowledgements**

We are grateful to the Institute of Biophysics SB RAS, Laboratory of Controlled Biosynthesis of Phototrophic Organisms, for allowing the main author to be part time employed on relevant research job during many years and have a lot of free time to study & research Terraforming & related topics without which this work won't be possible, and for useful discussions about biosphere dynamics & Terraforming, especially Prof. Dr. Sergey Trifonov.